\documentclass[letterpaper, 10 pt, conference]{ieeeconf}  

\IEEEoverridecommandlockouts                              
\usepackage{graphics} 
\usepackage{times} 
\usepackage{amsmath, amsfonts} 
\usepackage{amssymb}  
\usepackage{url}
\usepackage{cite}
\usepackage{hyperref}

\usepackage{algorithmic}
\usepackage{graphicx}
\usepackage{textcomp}
\usepackage{xcolor}

\title{\LARGE \bf
Parameter Estimation-Based States Reconstruction of Uncertain Linear Systems with Overparameterization and Unknown Additive Perturbations}

\author{Anton Glushchenko, \textit{Member, IEEE} and Konstantin Lastochkin
\thanks{Financial support is in part provided by the Grants Council of the President of the Russian Federation (MD-1787.2022.4).}
\thanks{A. Glushchenko is with V.A. Trapeznikov Institute of Control Sciences of Russian Academy of Sciences, Moscow, Russia
        {\tt\small aiglush@ipu.ru}}%
\thanks{K. Lastochkin is with V.A. Trapeznikov Institute of Control Sciences of Russian Academy of Sciences, Moscow, Russia 
        {\tt\small lastconst@ipu.ru}}%
}

\begin{document}

\maketitle
\thispagestyle{empty}
\pagestyle{empty}

\begin{abstract}
The problem of state reconstruction is considered for uncertain linear time-invariant systems with overparameterization, arbitrary state-space matrices and unknown additive perturbation described by an exosystem. A novel adaptive observer is proposed to solve it, which, unlike known solutions, simultaneously: ({\emph{i}}) reconstructs the physical state of the original system rather than the virtual state of its observer canonical form, ({\emph{ii}}) ensures exponential convergence of the reconstruction error to zero when the condition of finite excitation is satisfied, ({\emph{iii}}) is applicable to systems, in which mentioned perturbation is generated by an exosystem with fully uncertain constant parameters. The proposed solution uses a recently published parametrization of uncertain linear systems with unknown additive perturbations, the dynamic regressor extension and mixing procedure, as well as a method of physical states reconstruction developed by the authors. Detailed analysis for stability and convergence has been provided along with simulation results to validate the theoretical analysis.\end{abstract}

\section{Introduction}
Simultaneous reconstruction of unmeasured system states and unknown system parameters is provided with the help of adaptive observers. Today in such a long-standing research area, which dates back to the well-known studies of R. Carroll \cite{b1} and K. S. Narendra \cite{b2}, two main design principles are presented \cite{b3} for linear time-invariant (LTI) systems.

Techniques of the first type \cite{b4, b5} are based on the well-known KYP lemma, which allows one (thanks to $PD = \linebreak = {C^{\rm{T}}}$) to substitute an unmeasured signal $\left( {{{\hat x}^{\rm{T}}} - {x^{\rm{T}}}} \right)PD$ with measurable output estimation error $\hat y - y$ in the equation for derivative of a conventional adaptive control Lyapunov function so as to obtain an implementable adaptative law. The drawback of aforementioned approaches is that they are applicable only to systems, which satisfy matching or extended matching output conditions (particular system transfer function from system uncertainty to measured \cite{b4} or estimated \cite{b5} output must have a relative degree one).

The techniques of second type \cite{b1, b2, b6, b7, b8, b9, b10, b11, b12} are based on the system transformation with the help of substitution $\xi = Tx$ to an observer canonical form, which provides dependence of uncertainty not from all states $x$, but only from the measurable output $y$ and input $u$. For systems represented in the observer canonical form many parametrizations (e.g. \cite{b6}, \cite{b7}, \cite{b11}) can be applied to obtain a linear regression equation (LRE) with respect to (w.r.t.) unknown parameters that allows one to use the well-known gradient descent \cite{b6}, recursive least squares \cite{b6} or Dynamic Regressor Extension and Mixing (DREM) \cite{b12} estimators with certainty equivalence Luenberger-like \cite{b1, b2, b6} or algebraic \cite{b7, b8, b9, b10, b11, b12} state observers. Main drawbacks of such techniques are that: \linebreak ({\emph{i}}) some excitation conditions are necessary to be met to ensure convergence of states and parameters estimates to their true values, and ({\emph{ii}}) only the estimates of the virtual states of the observer canonical form can be obtained (smooth identification of unknown invert transform ${T^{ - 1}}$ is a not a clear task \cite{b13}).

Recently some efforts to solve the problem ({\emph{ii}}) have been made, and as a result a novel adaptive observer \cite{b13} was proposed, which ensures reconstruction of physical (original) system states $x$ instead of virtual ones $\xi$ for linear systems overparameterized with physical system parameters. Main idea of the approach \cite{b13} is to obtain the LRE w.r.t. ${T_I} = \linebreak = {T^{ - 1}}$ with the help of some clever/tedious parametrization and use the estimates ${\hat T_I}$ to reconstruct the unmeasured states via inversion $\hat x = {\hat T_I}\hat \xi$. Further the result of \cite{b13} was extended in \cite{b14} to systems with a disturbance generated by an exosystem with known parameters and unknown initial conditions.

On the other hand, a very interesting adaptive algebraic observer for uncertain linear time-varying (LTV) systems with unknown additive perturbations has been proposed recently \cite{b12}. In comparison with \cite{b13, b14} the system is represented in the observer canonical form (so only the virtual states can be reconstructed), which parameters are formed by an exosystem with known parameters and unknown initial conditions, but at the same time the disturbance is generated by an exosystem with unknown parameters and unknown initial conditions. So, it is worth studying how the approach from \cite{b13, b14} can be combined with \cite{b12} to obtain the estimates of physical (not virtual) states of the system that is not represented in the observer canonical form. Therefore, contribution of this paper lies in extension of the results of \cite{b13, b14} to the class of systems with additive perturbation generated by an exosystem with fully uncertain time-invariant parameters with the help of parametrizations from \cite{b12}.

\textbf{Notation and Definitions.} Further the following notation is used: $\left| . \right|$ is the absolute value, $\left\| . \right\|$ is the suitable norm of $(.)$, ${I_{n \times n}}=I_{n}$ is an identity $n \times n$ matrix, ${0_{n \times n}}$ is a zero $n \times n$ matrix, $0_{n}$ stands for a zero vector of length $n$, $e_{i}$ is the $i^{th}$ unit vector from the Euclidean $n$-space basis, ${\rm{det}}\{.\}$ stands for a matrix determinant, ${\rm{adj}}\{.\}$ represents an adjoint matrix, a column, diagonal matrix and block diagonal matrix are denoted as ${\rm{col}}\left\{.\right\}$, ${\rm{diag}}\left\{.\right\}$ and ${\rm{bd}}\left\{.\right\}$. We also use the fact that for all (possibly singular) ${n \times n}$ matrices $M$ the following holds: ${\rm{adj}} \{M\} M  = {\rm{det}} \{M\}I_{n \times n}.$ For a mapping ${\mathcal F}{\rm{:\;}}{\mathbb{R}^n} \mapsto {\mathbb{R}^n}$  we denote its Jacobian by $\nabla_{x} {\mathcal F}\left( x \right) = \linebreak = {\textstyle{{\partial {\mathcal F}} \over {\partial x}}}\left( x \right)$. 

The below-given definitions of the heterogeneous mapping and the regressor finite excitation condition are used throughout the paper.

{\it \bf Definition 1.} \emph{A mapping ${\cal F}{\rm{:\;}}{\mathbb{R}^{{n_x}}} \mapsto {\mathbb{R}^{{n_{\cal F}} \times {m_{\cal F}}}}$ is heterogeneous of degree ${\ell _{\cal F}} \ge 1$ if there exists ${\Xi _{\cal F}}\left( {\omega} \right)=\linebreak = {\overline \Xi _{\cal F}}\left( {\omega} \right)\omega \left( t \right) \in {\mathbb{R}^{{\Delta _{\cal F}} \times {n_x}}}{\rm{,\;}}{\Pi _{\cal F}}\left( {\omega} \right) \in {\mathbb{R}^{{n_{\cal F}} \times {n_{\cal F}}}}$, and a mapping ${{\cal T}_{\cal F}}{\rm{:\;}}{\mathbb{R}^{{\Delta _{\cal F}}}} \mapsto {\mathbb{R}^{{n_{\cal F}} \times {m_{\cal F}}}}$ such that for all $ \omega\left(t\right) \in \mathbb{R}$ and $x \in {\mathbb{R}^{{n_x}}}$ the following conditions hold:}
\begin{equation}\label{eq1}
\begin{array}{c}
{\Pi _{\cal F}}\left( {\omega} \right){\cal F}\left( x \right) = {{\cal T}_{\cal F}}\left( {{\Xi _{\cal F}}\left( {\omega} \right)x} \right){\rm{, }}\\
{\rm{det}}\left\{ {{\Pi _{\cal F}}\left( {\omega} \right)} \right\} \ge {\omega ^{{\ell _{_{\cal F}}}}}\left( t \right)\!{\rm{,\;}}\\{\Xi _{\cal F}}_{ij}\!\left( {\omega} \right) = {c_{ij}}{\omega ^{\ell_{ij}} }\left( t \right)\!{\rm{,\;}}{{\overline \Xi }_{{\cal F}ij}}\left( \omega  \right) = {c_{ij}}{\omega ^{{\ell _{ij}} - 1}}{\rm{,}}\\
{c_{ij}} \in \left\{ {0,{\rm{ 1}}} \right\}{\rm{,\;}}{\ell _\mathcal{F}} \geqslant 1,{\text{\;}}{\ell_{ij}}  \geqslant 1.
\end{array}
\end{equation}

For instance, for the mapping ${\cal F}\left( x \right) = {\rm{col}}\left\{ {{x_1}{x_2}{\rm{,\;}}{x_1}} \right\}$ with ${\Pi _{\cal F}}\left( \omega  \right) = {\rm{diag}}\left\{ {{\omega ^2}{\rm{,\;}}\omega } \right\}{\rm{,\;}}{\Xi _{\cal F}}\left( \omega  \right) = \omega{I_2},\;{\overline \Xi _{\cal F}}\left( \omega  \right) = {I_2}$ \linebreak we have that ${{\cal T}_{\cal F}}\left( {{\Xi _{\cal F}}\left( \omega  \right)x} \right) = {{\cal T}_{\cal F}}\left( {{{\overline \Xi }_{\cal F}}\left( \omega  \right)\omega x} \right) =\linebreak={\rm{col}}\left\{ {\omega {x_1}\omega {x_2}{\rm{,\;}}\omega {x_1}} \right\}$.

Definition 1 is useful for the following two tasks. 

{\emph{Task 1}}. Let us suppose that we need to measure (compute) the multiplication ${\Pi _{\cal F}}\left( \omega  \right){\cal F}\left( x \right)$, where ${\Pi _{\cal F}}\left( \omega  \right)$ is a measurable/computable signal and ${\cal F}\left( x \right)$ is an unknown. Then, if we have a measurable signal ${\cal Y}\left( t \right) = \omega \left( t \right)x$  with $\omega \left( t \right) \in \mathbb{R}$, then, owing to \eqref{eq1}, the solution to Task 1 is a measurable signal ${{\cal T}_{\cal F}}\left( {{{\overline \Xi }_{\cal F}}\left( \omega  \right){\cal Y}} \right)$. For instance, for the example under consideration ${{\cal T}_{\cal F}}\left( {{{\overline \Xi }_{\cal F}}\left( \omega  \right){\cal Y}} \right) = {\rm{col}}\left\{ {{{\cal Y}_1}{{\cal Y}_2}{\rm{,\;}}{{\cal Y}_1}} \right\} = \linebreak = {\Pi _{\cal F}}\left( \omega  \right){\cal F}\left( x \right)$. 

{\emph{Task 2}}. Let us suppose that an identification problem $\mathop {{\rm{lim}}}\limits_{t \to \infty } \left\| {\hat {\cal F}\left( t \right) - {\cal F}\left( x \right)} \right\| = 0$ is to be solved in case we have only ${\cal Y}\left( t \right) = \omega \left( t \right)x$ with $\omega \left( t \right) \in \mathbb{R}$. Also it is forbidden to obtain the estimate of $\hat x\left( t \right)$ and apply the certainty equivalence substitution $\hat {\cal F}\left( t \right) = {\cal F}\left( {\hat x} \right)$ as the Lipshitz condition $\left\| {{\cal F}\left( {\hat x} \right) - {\cal F}\left( x \right)} \right\| \le L\left\| {\hat x - x} \right\|$ can be violated for ${\cal F}\left( x \right)$, and consequently the convergence of the argument error $\hat x - x$ does not imply the convergence of the functional error ${\cal F}\left( {\hat x} \right) - {\cal F}\left( x \right)$. Then, if ${\cal F}\left( x \right)$ is heterogeneous in the sense of \eqref{eq1}, the adaptive law for $\hat {\cal F}\left( t \right)$ can be derived on the basis of LRE ${{\cal T}_{\cal F}}\left( {{{\overline \Xi }_{\cal F}}\left( \omega  \right){\cal Y}} \right) = {\Pi _{\cal F}}\left( \omega  \right){\cal F}\left( x \right)$ with both measurable regressand ${{\cal T}_{\cal F}}\left( {{{\overline \Xi }_{\cal F}}\left( \omega  \right){\cal Y}} \right)$ and regressor ${\Pi _{\cal F}}\left( \omega  \right)$. It should be mentioned that the regression equation ${\cal Y}\left( t \right) = \omega \left( t \right)x$ with a scalar regressor $\omega \left( t \right)$ can be obtained from an arbitrary regression $y\left( t \right) = {\varphi ^{\rm{T}}}\left( t \right)x$, $\varphi \left( t \right) \in {\mathbb{R}^{{n_x}}}$ by application of a well-known DREM procedure \cite{b12}.

{\it \bf Definition 2.} \emph{A regressor $\varphi \left( t \right) \in {\mathbb{R}^n}$ is finitely exciting $(\varphi \in {\rm{FE}})$  over the time range $\left[ {t_r^ + {\rm{,\;}}{t_e}} \right]$  if there exists $t_r^ +  \ge 0$, ${t_e} > t_r^ +$ and $\alpha$ such that the following inequality holds:}
\begin{equation}\label{eq2}
\int\limits_{t_r^ + }^{{t_e}} {\varphi \left( \tau  \right){\varphi ^{\rm{T}}}\left( \tau  \right)d} \tau  \ge \alpha {I_n}{\rm{,}}
\end{equation}
\emph{where $\alpha > 0$ is an excitation level, $I_{n}$ is an identity matrix.}

The inequality \eqref{eq2} is a necessary and sufficient condition for identifiability of unknown parameters $x \in {\mathbb{R}^n}$ of a regression equation $y\left( t \right) = {\varphi ^{\rm{T}}}\left( t \right)x$ \cite{b15}.

\section{Problem Statement}

A class of uncertain linear time-invariant overparametrized systems affected by bounded external disturbances are considered\footnote{Dependencies from $\theta$ and $t$ can be further suppressed for the sake of brevity.}:
\begin{equation}\label{eq3}
\begin{array}{l}
\dot x\left( t \right) = A\left( \theta  \right)x\left( t \right) + B\left( \theta  \right)u\left( t \right) + D\left( \theta  \right)\delta \left( t \right){\rm{,}}\\
y\left( t \right) = {C^{\rm{T}}}x\left( t \right){\rm{,\;}}x\left( {{t_0}} \right) = {x_0}{\rm{,}}
\end{array}
\end{equation}
where $x\left( t \right) \in {\mathbb{R}^n}$ is the original (physical) system states with unknown initial conditions ${x_0}$, $\delta \left( t \right) \in \mathbb{R}$ is a bounded external disturbance, $C \in {\mathbb{R}^n}$ denotes a known vector,  $A{\rm{:\;}}{\mathbb{R}^{{n_\theta }}} \mapsto {\mathbb{R}^{n \times n}}{\rm{,}}$ $B{\rm{:\;}}{\mathbb{R}^{{n_\theta }}} \!\mapsto\! {\mathbb{R}^n}{\rm{,\;}}$ $D{\rm{:\;}}{\mathbb{R}^{{n_\theta }}} \mapsto {\mathbb{R}^n}$ denote known mappings with unknown parameter $\theta  \in {\mathbb{R}^{{n_\theta }}}$. The pair $\left( {{C^{\rm{T}}}{\rm{,\;}}A\left( \theta  \right)} \right)$ is completely observable with $\theta \in D_{\theta}$ and only control $u\left( t \right) \in \mathbb{R}$ and output $y\left( t \right) \in \mathbb{R}$ signals are measurable.

Considering the control signal, disturbances and structure of the system, the following assumptions are adopted.

{\it \bf Assumption 1.} \emph{For all $t \ge {t_0}$ the control signal $u\left( t \right)$ ensures existence and boundedness of trajectories of the system \eqref{eq3}.}

{\it \bf Assumption 2.} \emph{Relative degree ${r_\delta }$ of the transfer function from $\delta \left( t \right)$ to $y\left( t \right)$ is equal to $n$, i.e.}
\begin{equation*}
\begin{array}{c}
{C^{\rm{T}}}D\left( \theta  \right) \!=\! {C^{\rm{T}}}A\left( \theta  \right)D\left( \theta  \right) \!=\!  \ldots  \!=\! {C^{\rm{T}}}{A^{n - 2}}\left( \theta  \right)D\left( \theta  \right) = 0,\\
{\rm{ }}{C^{\rm{T}}}{A^{n - 1}}\left( \theta  \right)D\left( \theta  \right) \ne 0.
\end{array}
\end{equation*}

{\it \bf Assumption 3.} \emph{The disturbance $\delta \left( t \right)$ is generated by:
\begin{equation}\label{eq4}
\begin{array}{l}
{{\dot x}_\delta }\left( t \right) = {{\cal A}_\delta }\left( \rho  \right){x_\delta }\left( t \right){\rm{,\;}}{x_\delta }\left( {{t_0}} \right) = {x_{\delta 0}}{\rm{,}}\\
\delta \left( t \right) = h_\delta ^{\rm{T}}{x_\delta }\left( t \right){\rm{,}}
\end{array}
\end{equation}
where it is assumed that the pair $\left( {h_\delta ^{\rm{T}}{\rm{,\;}}{{\cal A}_\delta }\left( \rho  \right)} \right)$ is observable, the vectors ${h_\delta } \in {\mathbb{R}^{{n_\delta }}}{\rm{,\;}}\rho  \in {\mathbb{R}^{{n_\rho }}}$ are unknown and state vector ${x_\delta }\left( t \right) \in {\mathbb{R}^{{n_\delta }}}$ is unmeasurable, but the mapping ${{\cal A}_\delta }{\rm{:\;}}{\mathbb{R}^{{n_\rho }}} \mapsto {\mathbb{R}^{{n_\delta } \times {n_\delta }}}$ is known and such that the eigenvalues of ${{\cal A}_\delta }\left( \rho  \right)$ have zero real parts for $\rho  \in {D_\rho }$.}

{\it \bf Assumption 4.} \emph{The parameters $\theta$ are globally structurally identifiable i.e. for almost any $\underline\theta\in D_{\theta}$, the following hold}
\begin{gather*}
\begin{array}{c}
    y\left(t,\;\underline{\theta},\; u\right)=y\left(t,\;\theta,\;u\right),\\\forall t \ge t_{0},\;\forall u\left(t\right) \in \mathbb{R}
\end{array}
\end{gather*}
{\emph{only for $\underline{\theta}=\theta.$}}

The goal is to design an adaptive observer of the system physical states, which guarantees that:
\begin{equation}\label{eq5}
\mathop {{\rm{lim}}}\limits_{t \to \infty } \left\| {\tilde x\left( t \right)} \right\| = 0{\rm{\;}}\left( {\exp } \right){\rm{,}}
\end{equation}
where $\tilde x\left( t \right) = \hat x\left( t \right) - x\left( t \right)$ is a state observation error of the system \eqref{eq3}, $\left(\rm{exp}\right)$ is an abbreviation for exponential rate of convergence. 

\section{Main Result}
The solution of problem \eqref{eq5} is proposed to be obtained by combining the results of \cite{b12} and \cite{b14}. In the first step the model \eqref{eq3} is represented into the observer canonical form with the help of transformation $\xi \left( t \right) = T\left( \theta  \right)x\left( t \right)$. Then a modified procedure from \cite{b12} is applied to solve the problem of virtual states $\xi \left( t \right)$ reconstruction. After that, using parametrizations from \cite{b14}, the transformation matrix ${T_I}\left( \theta  \right) = {T^{ - 1}}\left( \theta  \right)$ is identified and the system physical states are reconstructed via equation $\hat x\left( t \right) = {\hat T_I}\left( t \right)\hat \xi \left( t \right)$. The key feature of the proposed solution is that the estimate ${\hat T_I}\left( t \right)$ is obtained without identification of the parameters $\theta$ and implementation of recalculations ${\hat T_I}\left( t \right) = {T_I}\left( {\hat \theta } \right)$.

In accordance with the results from \cite{b1}, for each completely observable linear system \eqref{eq3} for all $\theta  \in {D_\theta }$ there exist nonsingular matrices:
\begin{gather*}
{\small{
\begin{array}{c}
{T_{I}}\left( \theta  \right) \!=\! {\begin{bmatrix}
{{A^{n - 1}}\left( \theta  \right){{\cal O}_n}\left( \theta  \right)}&{{A^{n - 2}}\left( \theta  \right){{\cal O}_n}\left( \theta  \right)}& \cdots &{{{\cal O}_n}\left( \theta  \right)}
\end{bmatrix}}{\rm{,}}\\
{{\cal O}_n}\left( \theta  \right) = {\cal O}\left( \theta  \right){{\begin{bmatrix}
{{0_{1 \times \left( {n - 1} \right)}}}&1
\end{bmatrix}}^{\rm{T}}}{\rm{,}}\\
{{\cal O}^{ - 1}}\!\left( \theta  \right) \!=\! {{\begin{bmatrix}
{C}&{{{\left( {A\left( \theta  \right)} \right)}^{\rm{T}}}C}& \!\cdots\! &{{{\left( {{A^{n - 1}}\left( \theta  \right)} \right)}^{\rm{T}}}C}
\end{bmatrix}}^{\rm{T}}}{\rm{,}}
\end{array}}}
\end{gather*}
which define the similarity transformation $\xi \left( t \right) = T\left( \theta  \right)x\left( t \right)$ to rewrite the system \eqref{eq3} in the observer canonical form:
\begin{equation}\label{eq7}
\begin{array}{l}
\dot \xi \left( t \right)\! =\! {A_0}\xi \left( t \right) + {\psi _a}\left( \theta  \right)y\left( t \right) +\;\;\;\;\;\;\;\;\;\;\\
\;\;\;\;\;\;\;\;\;\;\;\;\;\;\;\;\;\; + {\psi _b}\left( \theta  \right)u\left( t \right)+e_{n}{\psi _d}\left( \theta  \right)\delta \left( t \right){\rm{,}}
\end{array}
\end{equation}
\begin{equation}\label{eq8}
y\left( t \right) = C_0^{\rm{T}}\xi \left( t \right){\rm{,\;}}\xi \left( {{t_0}} \right) = {\xi _0}\left( \theta  \right) = T\left( \theta  \right){x_0}{\rm{,}}
\end{equation}
where
\begin{gather*}
\begin{array}{c}
{\psi _a}\left( \theta  \right) = T\left( \theta  \right)A\left( \theta  \right){T^{ - 1}}\left( \theta  \right){C_0}{\rm{,\;}}{\psi _b}\left( \theta  \right) = T\left( \theta  \right)B\left( \theta  \right){\rm{,}}\\
e_{n}{\psi _d}\left( \theta  \right) = T\left( \theta  \right)D\left( \theta  \right){\rm{,}}\\
{A_0} = {\begin{bmatrix}
{{0_n}}&{\begin{array}{*{20}{c}}
{{I_{n - 1}}}\\
{{0_{1 \times \left( {n - 1} \right)}}}
\end{array}}
\end{bmatrix}}{\rm{,\;}}\begin{array}{*{20}{c}}
{C_0^{\rm{T}} = {C^{\rm{T}}}{T^{ - 1}}\left( \theta  \right) = }\\
{ = {\begin{bmatrix}
1&{0_{n - 1}^{\rm{T}}}
\end{bmatrix}} }
\end{array}{\rm{,}}
\end{array}
\end{gather*}
${T_{I}}\left( \theta  \right){\rm{:}}=T^{-1}\left( \theta  \right)$, ${{\cal O}_n}$ is the $n^{th}$ column of the matrix that is inverse to ${{\cal O}^{ - 1}}\left( \theta  \right)$, 
$e_{n}$ is the $n^{th}$ unit vector from Euclidean space basis,
$\xi \left( t \right) \in {\mathbb{R}^n}$ denotes state vector of the observer canonical form with unknown initial conditions ${\xi _0}$, the vector ${C_0} \in {\mathbb{R}^n}$ and mappings ${\psi_a}{\rm{,\;}}{\psi _b}{\rm{,\;}}e_{n}{\psi _d}{\rm{:\;}}{D_\theta } \mapsto {D_\psi }$ are known. The equality $T\left( \theta  \right)D\left( \theta  \right) = {e_n}{\psi _d}\left( \theta  \right)$ is satisfied owing to the assumption 2.

With some straightforward modification of the results from \cite{b12}, the following is claimed for the state and unknown parameters of the system defined by equations \eqref{eq7} and \eqref{eq8}.

{\bf{Proposition 1.}} \emph{There exists sufficiently large ${t_{\epsilon}} \ge {t_0}$ such that for all $t \ge {t_{\epsilon}}$ the following claims hold:}
\begin{equation} \label{eq9}
    \overline q\left( t \right) = {f^{\rm{T}}}F\left( t \right) + y\left( t \right) - C_0^{\rm{T}}z\left( t \right) = \overline \varphi _e^{\rm{T}}\left( t \right){\eta _e}\left( \psi  \right){\rm{,}}
\end{equation}
\begin{equation}\label{eq10}
\begin{array}{l}
\xi \left( t \right) =
  z\left( t \right) + \Omega \left( t \right){\psi _a}\left( \theta  \right) + P\left( t \right){\psi _b}\left( \theta  \right)+\\+{\cal O}_e^{ - 1}{{\cal O}_\Gamma }\left( \Gamma  \right)\left( {F\left( t \right) - N\left( t \right){\psi _a}\left( \theta  \right) - H\left( t \right){\psi _b}\left( \theta  \right)} \right) {\rm{,}}
\end{array}
\end{equation}
\emph{where}
\begin{equation}\label{eq11}
\begin{array}{c}
{{\overline \varphi }_e}\left( t \right) = {\begin{bmatrix}
{{\Omega ^{\rm{T}}}{C_0} + {N^{\rm{T}}}f}\\
{{P^{\rm{T}}}{C_0} + {H^{\rm{T}}}f}\\
F\\
{vec\left( N \right)}\\
{vec\left( H \right)}
\end{bmatrix}}{\rm{,\;}}{\eta _e}\left( \psi  \right) = {\begin{bmatrix}
{{\psi _a}}\\
{{\psi _b}}\\
\Gamma \\
{ - {\psi _a} \otimes \Gamma }\\
{ - {\psi _b} \otimes \Gamma }
\end{bmatrix}} {\rm{,}}\\
\psi  = {{\begin{bmatrix}
{\psi _a^{\rm{T}}\left( \theta  \right)}&{\psi _b^{\rm{T}}\left( \theta  \right)}&{{\Gamma ^{\rm{T}}}}
\end{bmatrix}}^{\rm{T}}}{\rm{,}}
\end{array}
\end{equation}
\begin{equation} \label{eq12}
   {{\cal O}_e} = {\begin{bmatrix}
{C_0^{\rm{T}}}\\
{C_0^{\rm{T}}{A_K}}\\
 \vdots \\
{C_0^{\rm{T}}A_K^{n - 1}}
\end{bmatrix}} {\rm{,\;}}{{\cal O}_\Gamma }\left( \Gamma  \right) = {\begin{bmatrix}
{{{\left( {\Gamma  - f} \right)}^{\rm{T}}}}\\
{{{\left( {\Gamma  - f} \right)}^{\rm{T}}}{A_\Gamma }}\\
 \vdots \\
{{{\left( {\Gamma  - f} \right)}^{\rm{T}}}A_\Gamma ^{n - 1}}
\end{bmatrix}}{\rm{,}}
\end{equation}
\begin{equation} \label{eq13}
    \begin{array}{c}
\dot z\left( t \right) = {A_K}z\left( t \right) + Ky\left( t \right){\rm{,\;}}z\left( {{t_0}} \right) = {0_n}{\rm{,}}\\
\dot P\left( t \right) = {A_K}P\left( t \right) + {I_n}u\left( t \right){\rm{,\;}}P\left( {{t_0}} \right) = {0_{n \times n}}{\rm{,}}\\
\dot \Omega \left( t \right) = {A_K}\Omega \left( t \right) + {I_n}y\left( t \right){\rm{,\;}}\Omega \left( {{t_0}} \right) = {0_{n \times n}}{\rm{,}}\\
\dot F\left( t \right) = {A_f}F\left( t \right) + \hfill \\ \hfill  +{e_n}\left[ {y\left( t \right) - C_0^{\rm{T}}z\left( t \right)} \right]{\rm{,\;}}F\left( {{t_0}} \right) = {0_n}{\rm{,}}\\
\dot H\left( t \right) = {A_f}H\left( t \right) + {e_n}C_0^{\rm{T}}P\left( t \right){\rm{,\;}}H\left( {{t_0}} \right) = {0_{n \times n}},\\
\dot N\left( t \right) = {A_f}N\left( t \right) + {e_n}C_0^{\rm{T}}\Omega \left( t \right){\rm{,\;}}N\left( {{t_0}} \right) = {0_{n \times n}},
\end{array}
\end{equation}

\emph{Here the Hurwitz matrices ${A_K}{\rm{,\;}}{A_\Gamma }{\rm{,\;}}{A_f}$ have the following structure:}
\begin{equation*}
\begin{array}{c}
    {A_K} = {\begin{bmatrix}
{ - K}&{\begin{matrix}
{{I_{n - 1}}}\\
{{0_{1 \times \left( {n - 1} \right)}}}
\end{matrix}}
\end{bmatrix}}{\rm{,\;}}{A_\Gamma } = {\begin{bmatrix}
{\begin{matrix}
{{0_{\left( {n - 1} \right) \times 1}}}&{{I_{n - 1}}}
\end{matrix}}\\
{{\Gamma ^{\rm{T}}}}
\end{bmatrix}}{\rm{, }}\\{A_f} = {\begin{bmatrix}
{\begin{matrix}
{{0_{\left( {n - 1} \right) \times 1}}}&{{I_{n - 1}}}
\end{matrix}}\\
{{f^{\rm{T}}}}
\end{bmatrix}}{\rm{,}}
\end{array}
\end{equation*}
{\emph{while it holds that $\sigma \left( {{A_\Gamma }} \right){\rm{:}} = \sigma \left( {{{\cal A}_\delta }\left( \rho  \right)} \right) \cup {0_{n - {n_\delta }}}$.}}

{\emph{Proof of the proposition is given in \cite{b12} and, using alternative notation for the considered time-invariant system, it is also presented in \cite{b16}.}}

~

With the help of equation \eqref{eq9} we can obtain the estimate $\hat \psi \left( t \right)$ and substitute it into the second equation to have the virtual state estimate $\hat \xi \left( t \right)$ \cite{b12}. However, in the above derived parameterization \eqref{eq9} the dimension of regressor ${\overline \varphi _e}\left( t \right)$ is equal to $3n + 2{n^2}$. Consequently, it is complicated or even impossible exercise to ensure its {\emph{bona fide}} finite excitation even for the case $n = 1$ (for validation of this thesis please see the Supplementary material \cite{b16}). To overcome this obstacle, it is necessary to use some kind of {\emph{a priori}} knowledge about the regressor and unknown parameters, e.g. the equality $e_i^{\rm{T}}{\overline \varphi _e}\left( t \right) = e_j^{\rm{T}}{\overline \varphi _e}\left( t \right){\rm{,\;}}i \ne j$ for the regressor or $e_i^{\rm{T}}{\eta _e}\left( \psi \right) = 0$ for unknown parameters\footnote{Unfortunately, for now we cannot provide a theoretically sound proof that there always exist such $i$ and $j$, $i \ne j$, that $e_i^{\rm{T}}{\overline \varphi _e}\left( t \right) = e_j^{\rm{T}}{\overline \varphi _e}\left( t \right)$. However, in many simulation examples this is the case (see \cite{b16} and Section IV).} So, motivated by this observation, we adopt the following hypothesis about {\emph{reductionabilty}} of the unknown parameters ${\eta _e}\left( \psi \right)$ dimension.

{\bf{Hyphothesis 1.}} \emph{There exist known matrices \linebreak${{\cal D}_\eta } \in {\mathbb{R}^{\left( {3n + 2{n^2}} \right) \times {n_\eta }}}$, ${{\cal L}_\eta } \in {\mathbb{R}^{{n_\eta } \times \left( {3n + 2{n^2}} \right)}}{\rm{,\;}}{{\cal L}_\psi } \in {\mathbb{R}^{3n \times {n_\eta }}}$ and ${{\cal L}_{ab}} \in {\mathbb{R}^{{n_\theta } \times 3n}}$ such that ${n_\eta } < 3n + 2{n^2}$ and the following equations hold for all ${t_{\epsilon}} \ge {t_0}$:}
\begin{equation} \label{eq14}
\begin{array}{c}
\overline q\left( t \right) = {{\overline \varphi }^{\rm{T}}}\left( t \right)\eta \left( \psi  \right){\rm{,}}\\
{{\overline \varphi }^{\rm{T}}}\left( t \right) = \overline \varphi _e^{\rm{T}}\left( t \right){{\cal D}_\eta }{\rm{,\;}}\eta \left( \psi  \right) = {{\cal L}_\eta }{\eta _e}\left( \psi  \right){\rm{, }}
\end{array}
\end{equation}
\begin{equation} \label{eq15}
    \begin{array}{c}
{\rm{de}}{{\rm{t}}^2}\left\{ {{\nabla _\psi }{{\cal L}_\psi }\eta \left( \psi  \right)} \right\} > 0,{\rm{ }}\\
{\rm{de}}{{\rm{t}}^2}\left\{ {{\nabla _\theta }{\psi _{ab}}\left( \theta  \right)} \right\} > 0,{\rm{  }}{\psi _{ab}}\left( \theta  \right) = {{\cal L}_{ab}}\psi 
\end{array}
\end{equation}
{\emph{and moreover for the new regressor $\overline \varphi \left( t \right)$ the finite excitation condition is met (i.e. $\overline \varphi  \in {\rm{FE}}$) even if the old regressor ${\overline \varphi _e}\left( t \right)$ is not finitely exciting (i.e. ${\overline \varphi _e} \notin {\rm{FE}}$).}}

It should be mentioned that eliminators ${{\cal L}_\eta }$ and ${{\cal D}_\eta }$ always exist in the case $n > {n_\delta }$ as vector $\Gamma$ necessarily has $n - {n_\delta }$ zeros elements (as eigenvalues of ${{\cal A}_\delta }\left( \rho  \right)$ have zero real part by Assumption 3), which provide the well-understandable eliminations from vectors $\Gamma $ and $- {\psi _a} \otimes \Gamma {\rm{,\;}} - {\psi _b} \otimes \Gamma $. Moreover, hyphothesis is not restrictive and can be validated in online fashion via (\emph{i}) check whether we have $i$ and $j$, for which the conditions $i \ne j\;e_i^{\rm{T}}{\overline \varphi _e}\left( t \right) = e_j^{\rm{T}}{\overline \varphi _e}\left( t \right)$ hold for some time interval, (\emph{ii}) verification that the inequalities \eqref{eq15} have the solutions, (\emph{iii}) usage of \emph{ a priori} knowledge about zero elements of the vector ${\eta _e}\left( \psi \right)$ (for example, see equations \eqref{eq36} from numerical experiments).

Equation \eqref{eq14} from Hypothesis 1 (with $\overline \varphi  \in {\rm{FE}}$) is the condition that provides an opportunity to identify the parameters $\eta \left( \psi  \right)$ even if necessary condition for identifiability of ${\eta _e}\left( \psi  \right)$ is violated.

Equation \eqref{eq15} describes  sufficient conditions to reconstruct both the unknown parameters $\psi$ from $\eta \left( \psi  \right)$ and physical parameters $\theta$ from $\psi$. The requirements \eqref{eq15} are related to structural identifiability conditions, i.e. if the Assumption 4 is met then requirements from \eqref{eq15} are also met, and therefore there exists a function ${{\cal F}_\psi }{\rm{:\;}}{D_\eta } \mapsto {D_\psi } \cup {D_\rho }$ and inverse function ${{\cal F}_\theta }{\rm{:\;}}{D_\psi } \mapsto {D_\theta }$ such that:
\begin{equation} \label{eq16}
    \psi  = {{\cal F}_\psi }\left( \eta  \right){\rm{,\;}}\theta  = {{\cal F}_\theta }\left( {{\psi _{ab}}} \right).
\end{equation}

It should be mentioned that the function ${{\cal F}_\psi }\left( \eta  \right)$ includes elements of two types: ({\emph{i}}) to transform $\eta$ into zero elements of vector $\psi$ and (\emph{ii}) to obtain the remaining generally non-zero elements from non-zero elements of the vector $\eta$. The first part always exists and an existence condition for (\emph{ii}) is given in equation \eqref{eq15}.

So, when Hyphothesis 1 is met, the regression equation \eqref{eq9} from proved proposition 1 can be transformed into a set of scalar regression equations with a non-vanishing scalar regressor.

{\bf{Proposition 2.}} \emph{Using procedures of extension:}
\begin{equation} \label{eq17}
\begin{array}{c}
q\left( t \right) = \int\limits_{{t_{\epsilon}}}^t {{e^{ - \sigma \left( {\tau  - {t_{\epsilon}}} \right)}}\overline \varphi \left( \tau  \right)\overline q\left( \tau  \right)d\tau } {\rm{,}}\\
\varphi \left( t \right) = \int\limits_{{t_{\epsilon}}}^t {{e^{ - \sigma \left( {\tau  - {t_{\epsilon}}} \right)}}\overline \varphi \left( \tau  \right){{\overline \varphi }^{\rm{T}}}\left( \tau  \right)d\tau } {\rm{,}}\\
q\left( {{t_{\epsilon}}} \right) = {0_{{\rm{2}}n}}{\rm{,\;}}
\varphi \left( {{t_{\epsilon}}} \right) = {0_{{\rm{2}}n \times {\rm{2}}n}}{\rm{,}}
\end{array}
\end{equation}
{\emph{and mixing:}}
\begin{equation}\label{eq18}
\begin{matrix}
{{\cal Y}\left( t \right) = k\left( t \right) \cdot {\rm{adj}}\left\{ {\varphi \left( t \right)} \right\}q\left( t \right){\rm{,}}}\\{\Delta \left( t \right) = k\left( t \right) \cdot {\rm{det}}\left\{ {\varphi \left( t \right)} \right\}{\rm{,}}}
\end{matrix}
\end{equation}
{\emph{for all $t \ge {t_{\epsilon}}$ a set of scalar regression equations is obtained:}}
\begin{equation}\label{eq19}
{\cal Y}\left( t \right) = \Delta \left( t \right)\eta \left( \psi  \right),
\end{equation}
{\emph{where $\sigma  > 0$ is a damping factor, $k\left( t \right) \ge {k_{{\rm{min}}}} > 0$ is an amplitude amplifier and for all $t \ge {t_e}$ it holds that $\Delta \left( t \right) \ge \\ \ge {\Delta _{{\rm{min}}}} > 0$ when $\overline \varphi  \in {\rm{FE}}$ over $\left[ {{t_{\epsilon}}{\rm{,\;}}{t_e}} \right]$}.}

\emph{Proof of proposition is presented in Supplementary material \cite{b16}.}

~

So, having the regression equations \eqref{eq10}, \eqref{eq19} and inverse functions \eqref{eq16} at hand, we are in position to try to obtain the estimate $\hat \eta \left( t \right)$ with the help of gradient descent estimator:
\renewcommand{\theequation}{\arabic{equation}a}
\begin{equation}\label{eq20a}
    \dot {\hat \eta} \left( t \right) =  - \gamma \Delta \left( t \right)\left( {\Delta \left( t \right)\hat \eta \left( t \right) - {\cal Y}\left( t \right)} \right){\rm{,\;}}\hat \eta \left( {{t_0}} \right) = {\hat \eta _0}{\rm{,}}
\end{equation}
and use the following certainty equivalence observer to achieve the goal \eqref{eq5}:
\setcounter{equation}{18}
\renewcommand{\theequation}{\arabic{equation}b}
\begin{equation}\label{eq20b}
\begin{array}{l}
\hat \psi \left( t \right) = {{\cal F}_\psi }\left( {\hat \eta } \right) = {{\begin{bmatrix}
{\hat \psi _a^{\rm{T}}\left( t \right)}&{\hat \psi _b^{\rm{T}}\left( t \right)}&{{{\hat \Gamma }^{\rm{T}}}\left( t \right)}
\end{bmatrix}}^{\rm{T}}}{\rm{,}}\\
\hat \theta \left( t \right) = {{\cal F}_\theta }\left( {{{\cal L}_{ab}}\hat \psi } \right),\\
\hat x\left( t \right) = {T_I}\left( {\hat \theta } \right)\left[ {z + \Omega {{\hat \psi }_a} + P{{\hat \psi }_b} + } \right.\hfill\\\hfill
\left. { + {\cal O}_e^{ - 1}{{\cal O}_\Gamma }\left( {\hat \Gamma } \right)\left( {F - N{{\hat \psi }_a} - H{{\hat \psi }_b}} \right)} \right],
\end{array}
\end{equation}
where $\gamma  > 0$.
\renewcommand{\theequation}{\arabic{equation}}

However, the substitutions \eqref{eq20b} of dynamic estimates $\hat \eta \left( t \right){\rm{,\;}}\hat \psi \left( t \right)$ and $\hat \theta \left( t \right){\rm{,\;}}\hat \Gamma \left( t \right)$ into nonlinear mappings are free from the singularity issue iff the corresponding mappings satisfy the well-known global Lipschitz continuous condition, i.e. when there exists some constant $L > 0$ such that:
\begin{equation} \label{eq21}
    \left\| {{\begin{bmatrix}
{{T_I}\left( {\hat \theta } \right)}\\
{{{\cal O}_\Gamma }\left( {\hat \Gamma } \right)}\\
{{{\cal F}_\psi }\left( {\hat \eta } \right)}\\
{{{\cal F}_\theta }\left( {{{\cal L}_{ab}}\hat \psi } \right)}
\end{bmatrix}} - {\begin{bmatrix}
{{T_I}\left( \theta  \right)}\\
{{{\cal O}_\Gamma }\left( \Gamma  \right)}\\
{{{\cal F}_\psi }\left( \eta  \right)}\\
{{{\cal F}_\theta }\left( {{\psi _{ab}}} \right)}
\end{bmatrix}}} \right\| \le L\left\| {{\begin{bmatrix}
{\hat \theta }\\
{\hat \Gamma }\\
{\hat \eta }\\
{\hat \psi }
\end{bmatrix}} - {\begin{bmatrix}
\theta \\
\Gamma \\
\eta \\
\psi 
\end{bmatrix}}} \right\|
\end{equation}
for all values of parameters estimates $\left( {\widehat .} \right)$.

The inverse functions \eqref{eq16} and transformation matrix may trivially include the division operation, and consequently ${{\cal F}_\psi }\left( {\hat \eta } \right){\rm{,\;}}{{\cal F}_\theta }\left( {{{\cal L}_{ab}}\hat \psi } \right){\rm{,\;}}{T_I}\left( {\hat \theta } \right)$ are not singularity free operations, therefore, we do not have opportunity to substitute obtained dynamic estimates into such mappings. To overcome this second obstacle, we need to obtain the estimates of parameters $\psi {\rm{,\;}}\theta {\rm{,\;}}{T_I}\left( \theta  \right){\rm{,\;}}{{\cal O}_\Gamma }\left( \Gamma  \right)$ without using substitution operations, so the regression equations w.r.t. them are required to be parametrized. For this purpose we adopt the following second Hypothesis, which is motivated by the main results of studies \cite{b13, b14}.

{\bf{Hyphothesis 2.}} \emph{There exist heterogenous mappings ${{\cal G}_\psi }{\rm{:\;}}{\mathbb{R}^{3n}} \mapsto {\mathbb{R}^{3n \times 3n}}$, ${{\cal S}_\psi }{\rm{:\;}}{\mathbb{R}^{3n}} \mapsto {\mathbb{R}^{3n}}$, ${{\cal G}_\theta }{\rm{:\;}}{\mathbb{R}^{{n_\theta }}} \mapsto {\mathbb{R}^{{n_\theta } \times {n_\theta }}}$, ${{\cal S}_\theta }{\rm{:\;}}{\mathbb{R}^{{n_\theta }}} \mapsto {\mathbb{R}^{{n_\theta }}}$, ${\cal P}{\rm{:\;}}{\mathbb{R}^{{n_\theta }}} \mapsto {\mathbb{R}^{n \times n}}$, ${\cal Q}{\rm{:\;}}{\mathbb{R}^{{n_\theta }}} \mapsto {\mathbb{R}^n}$ such that:}
\begin{equation}\label{eq22}
    \begin{array}{c}
{{\cal S}_\psi }\left( \eta  \right) = {{\cal G}_\psi }\left( \eta  \right)\psi {\rm{,}}\\
{\Pi _\psi }\left( \Delta  \right){{\cal G}_\psi }\left( \eta  \right) = {{\cal T}_{{{\cal G}_\psi }}}\left( {{\Xi _{{{\cal G}_\psi }}}\left( \Delta  \right)\eta } \right){\rm{,}}\\
{\Pi _\psi }\left( \Delta  \right){{\cal S}_\psi }\left( \eta  \right) = {{\cal T}_{{{\cal S}_\psi }}}\left( {{\Xi _{{{\cal S}_\psi }}}\left( \Delta  \right)\eta } \right){\rm{,}}
\end{array}
\end{equation}
\begin{equation}\label{eq23}
    \begin{array}{c}
{{\cal S}_\theta }\left( {{\psi _{ab}}} \right) = {{\cal G}_\theta }\left( {{\psi _{ab}}} \right)\theta {\rm{,}}\\
{\Pi _\theta }\left( {{{\cal M}_\psi }} \right){{\cal G}_\theta }\left( {{\psi _{ab}}} \right) = {{\cal T}_{{{\cal G}_\theta }}}\left( {{\Xi _{{{\cal G}_\theta }}}\left( {{{\cal M}_\psi }} \right){\psi _{ab}}} \right){\rm{,}}\\
{\Pi _\theta }\left( {{{\cal M}_\psi }} \right){{\cal S}_\theta }\left( {{\psi _{ab}}} \right) = {{\cal T}_{{{\cal S}_\theta }}}\left( {{\Xi _{{{\cal S}_\theta }}}\left( {{{\cal M}_\psi }} \right){\psi _{ab}}} \right){\rm{,}}
\end{array}
\end{equation}
\begin{equation}\label{eq24}
    \begin{array}{c}
{\cal Q}\left( \theta  \right) = {\cal P}\left( \theta  \right){T_I}\left( \theta  \right){\rm{,}}\\
{\Pi _{{T_I}}}\left( {{{\cal M}_\theta }} \right){\cal Q}\left( \theta  \right) = {{\cal T}_{\cal Q}}\left( {{\Xi _{\cal Q}}\left( {{{\cal M}_\theta }} \right)\theta } \right){\rm{,}}\\
{\Pi _{{T_I}}}\left( {{{\cal M}_\theta }} \right){\cal P}\left( \theta  \right) = {{\cal T}_{\cal P}}\left( {{\Xi _{\cal P}}\left( {{{\cal M}_\theta }} \right)\theta } \right){\rm{,}}
\end{array}
\end{equation}
\begin{equation}\label{eq25}
    {\Pi _{{{\cal O}_\Gamma }}}\left( {{{\cal M}_\psi }} \right){{\cal O}_\Gamma }\left( \Gamma  \right) = {{\cal T}_{{{\cal O}_\Gamma }}}\left( {{\Xi _{{{\cal O}_\Gamma }}}\left( {{{\cal M}_\psi }} \right)\Gamma } \right){\rm{,}}
\end{equation}
\emph{where}
\begin{equation*}
\begin{array}{l}
{\rm{det}}\left\{ {\cal I} \right\} \ge {\rm{arg}}{\left\{ {\cal I} \right\}^{{\ell _{\left( . \right)}}}}{\rm{,}}\\{\cal I} \in \left\{ {{\Pi _\psi }\left( \Delta  \right){\rm{,\;}}{\Pi _\theta }\left( {{{\cal M}_\psi }} \right){\rm{,\;}}{\Pi _{{T_I}}}\left( {{{\cal M}_\theta }} \right){\rm{,\;}}{\Pi _{{{\cal O}_\Gamma }}}\left( {{{\cal M}_\psi }} \right)} \right\}{\rm{,}}\\
{\rm{det}}\left\{ {\cal J} \right\} \ne 0,{\rm{\;}}{\cal J} \in \left\{ {{{\cal G}_\psi }\left( \eta  \right){\rm{,\;}}{{\cal G}_\theta }\left( {{\psi _{ab}}} \right){\rm{,\;}}{\cal P}\left( \theta  \right)} \right\}{\rm{,}}
\end{array}
\end{equation*}
\emph{and all mappings are known.}

In fact, hypothesis 2 requires that the mappings ${{\cal F}_\psi }\left( \eta  \right){\rm{,\;}}{{\cal F}_\theta }\left( {{\psi _{ab}}} \right)$ and ${T_I}\left( \theta  \right){\rm{,\;}}{{\cal O}_\Gamma }\left( \Gamma  \right)$ can be decomposed into a numerator and a denominator, which are heterogeneous functions in the sense of definition 1 (as follows from definition \eqref{eq12}, the denominator of ${{\cal O}_\Gamma }\left( \Gamma  \right)$  always equals to an identity matrix). Owing to the property ${\Xi _{\left( . \right)}}\left( . \right) = {\overline \Xi _{\left( . \right)}}\left( . \right) \cdot \left( . \right)$ from definition 1, the fact that hypothesis 2 is met allows one to transform the regression equation \eqref{eq19} w.r.t. $\eta \left( \psi  \right)$ into the regression equation w.r.t. $\psi {\rm{,\;}}{{\cal O}_\Gamma }\left( \Gamma  \right){\rm{,\;}}\theta {\rm{,\;}}{T_I}\left( \theta  \right)$ (for more details please see comments after definition 1, the example in Section IV, supplementary material \cite{b16} and studies \cite{b13, b14}).

The procedure to transform the regression equation \eqref{eq19} into regression equations with respect to $\psi {\rm{,\;}}{{\cal O}_\Gamma }\left( \Gamma  \right)$ and ${T_I}\left( \theta  \right)$ is described in the following lemma.

{\bf{Lemma.}} \emph{For all $t \ge {t_{\epsilon}}$ the unknown parameters $\kappa  = \linebreak = { {\begin{bmatrix}
{{\psi ^{\rm{T}}}}&{ve{c^{\rm{T}}}\left( {{{\cal O}_\Gamma }\left( \Gamma  \right)} \right)}&{ve{c^{\rm{T}}}\left( {{T_I}\left( \theta  \right)} \right)}
\end{bmatrix}}^{\rm{T}}}$ satisfy the following regression equation:}
\begin{equation} \label{eq26}
    \begin{array}{c}
{{\cal Y}_\kappa }\left( t \right) = {{\cal M}_\kappa }\left( t \right)\kappa {\rm{,}}\\
{{\cal Y}_\kappa }\left( t \right) = {\rm{adj}}\left\{ {{\rm{bd}}\left\{ {{{\cal M}_\psi }\left( t \right){\rm{,\;}}{{\cal M}_{{{\cal O}_\Gamma }}}\left( t \right){\rm{,\;}}{{\cal M}_{{T_I}}}\left( t \right)} \right\}} \right\} \\
 \times \left[ {\begin{array}{*{20}{c}}
{{{\cal Y}_\psi }\left( t \right)}\\
{vec\left( {{{\cal Y}_{{{\cal O}_\Gamma }}}\left( t \right)} \right)}\\
{vec\left( {{{\cal Y}_{{T_I}}}\left( t \right)} \right)}
\end{array}} \right]{\rm{,}}\\
{{\cal M}_\kappa }\left( t \right) = {\rm{det}}\left\{ {{\rm{bd}}\left\{ {{{\cal M}_\psi }{I_{3n}}{\rm{,\;}}{{\cal M}_{{{\cal O}_\Gamma }}}{I_{{n^2}}}{\rm{,\;}}{{\cal M}_{{T_I}}}{I_{{n^2}}}} \right\}} \right\}
\end{array}
\end{equation}
\emph{where:}

\emph{1) the regressand and regressor of ${{\cal Y}_\psi }\left( t \right) = {{\cal M}_\psi }\left( t \right)\psi$ are defined as follows:}
\begin{equation*}
\begin{array}{c}
{{\cal Y}_\psi }\left( t \right) = {\rm{adj}}\left\{ {{{\cal T}_{{{\cal G}_\psi }}}\left( {{{\overline \Xi }_{{{\cal G}_\psi }}}\left( \Delta  \right){\cal Y}} \right)} \right\}{{\cal T}_{{{\cal S}_\psi }}}\left( {{{\overline \Xi }_{{{\cal S}_\psi }}}\left( \Delta  \right){\cal Y}} \right),\\
{{\cal M}_\psi }\left( t \right) = {\rm{det}}\left\{ {{{\cal T}_{{{\cal G}_\psi }}}\left( {{{\overline \Xi }_{{{\cal G}_\psi }}}\left( \Delta  \right){\cal Y}} \right)} \right\},
\end{array}
\end{equation*}

\emph{2) the regressand and regressor of ${{\cal Y}_{{{\cal O}_\Gamma }}}\left( t \right) = \linebreak ={{\cal M}_{{{\cal O}_\Gamma }}}\left( t \right){{\cal O}_\Gamma }\left( \Gamma  \right)$ are defined as follows:}
\begin{equation*}
\begin{array}{c}
{{\cal Y}_{{{\cal O}_\Gamma }}}\left( t \right) = {\rm{adj}}\left\{ {{\Pi _{{{\cal O}_\Gamma }}}\left( {{{\cal M}_\psi }} \right)} \right\}{{\cal T}_{{{\cal O}_\Gamma }}}\left( {{{\overline \Xi }_{{{\cal O}_\Gamma }}}\left( {{{\cal M}_\psi }} \right){{\cal Y}_\Gamma }} \right){\rm{,}}\\
{{\cal M}_{{{\cal O}_\Gamma }}}\left( t \right) = {\rm{det}}\left\{ {{\Pi _{\cal O}}\left( {{{\cal M}_\psi }} \right)} \right\}{\rm{,}}
\end{array}
\end{equation*}
\emph{where ${{\cal Y}_\Gamma }\left( t \right) = {{\cal L}_\Gamma }{{\cal Y}_\psi }\left( t \right)$, and ${{\cal L}_\Gamma }$  is such that ${{\cal L}_\Gamma }\psi  = \Gamma $.}

\emph{3) the regressand and regressor of ${{\cal Y}_{{T_I}}}\left( t \right) = \linebreak = {{\cal M}_{{T_I}}}\left( t \right){T_I}\left( \theta  \right)$, considering auxiliary equations:}
\begin{equation*}
\begin{array}{c}
{{\cal Y}_\theta }\left( t \right) = {\rm{adj}}\left\{ {{{\cal T}_{{{\cal G}_\theta }}}\left( {{{\overline \Xi }_{{{\cal G}_\theta }}}\left( {{{\cal M}_\psi }} \right){{\cal Y}_{ab}}} \right)} \right\}{{\cal T}_{{{\cal S}_\theta }}}\left( {{{\overline \Xi }_{{{\cal S}_\theta }}}\left( {{{\cal M}_\psi }} \right){{\cal Y}_{ab}}} \right) {\rm{,}}\\
{{\cal M}_\theta }\left( t \right) = {\rm{det}}\left\{ {{{\cal T}_{{{\cal G}_\theta }}}\left( {{{\overline \Xi }_{{{\cal G}_\theta }}}\left( {{{\cal M}_\psi }} \right){{\cal Y}_{ab}}} \right)} \right\}{\rm{,}}
\end{array}
\end{equation*}
\emph{are defined as follows:}
\begin{equation*}
\begin{array}{c}
{{\cal Y}_{{T_I}}}\left( t \right) = {\rm{adj}}\left\{ {{{\cal T}_{\cal P}}\left( {{{\overline \Xi }_{\cal P}}\left( {{{\cal M}_\theta }} \right){{\cal Y}_\theta }} \right)} \right\}{{\cal T}_{\cal Q}}\left( {{{\overline \Xi }_{\cal Q}}\left( {{{\cal M}_\theta }} \right){{\cal Y}_\theta }} \right){\rm{,}}\\
{{\cal M}_{{T_I}}}\left( t \right) = {\rm{det}}\left\{ {{{\cal T}_{\cal P}}\left( {{{\overline \Xi }_{\cal P}}\left( {{{\cal M}_\theta }} \right){{\cal Y}_\theta }} \right)} \right\},
\end{array}
\end{equation*}
\emph{where ${{\cal Y}_{ab}}\left( t \right) = {{\cal L}_{ab}}{{\cal Y}_\psi }\left( t \right)$, and ${{\cal L}_{ab}}$  is defined in \eqref{eq15}.}

\emph{If additionally $\overline \varphi  \in {\rm{FE}}$ over $\left[ {{t_{\epsilon}}{\rm{,\;}}{t_e}} \right]$, then for all $t \ge {t_e}$ the inequality $\left| {{{\cal M}_\kappa }\left( t \right)} \right| \ge \underline {{{\cal M}_\kappa }}  > 0$ holds.}	

\emph{Proof of lemma is presented in Supplementary \cite{b16}.}

~

Using the regression equations \eqref{eq10} and \eqref{eq26}, we are in position to implement the adaptive observer of system \eqref{eq3} state:
\begin{equation} \label{eq31}
    \begin{array}{l}
\dot {\hat{ \kappa}} \left( t \right) = \dot {\tilde {\kappa}}\left( t \right) = \hfill \\\hfill = - \gamma {{\cal M}_\kappa }\left( t \right)\left( {{{\cal M}_\kappa }\left( t \right)\hat \kappa \left( t \right) - {{\cal Y}_\kappa }\left( t \right)} \right){\rm{,\;}}\hat \kappa \left( {{t_0}} \right) = {{\hat \kappa }_0}{\rm{,}}\\
\hat x\left( t \right) = {{\hat T}_I}\left( t \right)\hat \xi \left( t \right){\rm{,}}\\
\hat \xi \left( t \right) = {\cal O}_e^{ - 1}{{\hat {\cal O}}_\Gamma }\left( {F - N{{\hat \psi }_a} - H{{\hat \psi }_b}} \right) + \hfill\\\hfill
 + z + \Omega {{\hat \psi }_a} + P{{\hat \psi }_b}{\rm{, }}
\end{array}
\end{equation}
where $\gamma  > 0$.

The following expressions hold for the state reconstruction errors $\tilde \xi \left( t \right)$ and $\tilde x\left( t \right)$ (for proof please see \cite{b16}):
\begin{equation}\label{eq32}
\begin{array}{l}
\tilde x\left( t \right)
 = {{\tilde T}_I}\left( t \right)\tilde \xi \left( t \right) + {T_I}\left( \theta  \right)\tilde \xi \left( t \right) + {{\tilde T}_I}\left( t \right)\xi \left( t \right){\rm{,}}\hfill\\\hfill
\tilde \xi \left( t \right)
 = {\cal O}_e^{ - 1}{{\tilde {\cal O}}_\Gamma }\left( { - N{{\tilde \psi }_a} - H{{\tilde \psi }_b}} \right) + \hfill\\\hfill 
 + {\cal O}_e^{ - 1}{{\tilde {\cal O}}_\Gamma }\left( {F - N{\psi _a}\left( \theta  \right) - H{\psi _b}\left( \theta  \right)} \right) + \\
 + \Omega {{\tilde \psi }_a} + P{{\tilde \psi }_b} + {\cal O}_e^{ - 1}{{\cal O}_\Gamma }\left( \Gamma  \right)\left( { - N{{\tilde \psi }_a} - H{{\tilde \psi }_b}} \right).\hfill
\end{array}
\end{equation}

On the other hand, if the condition $\overline \varphi  \in {\rm{FE}}$ is met over $\left[ {{t_{\epsilon}}{\rm{,\;}}{t_e}} \right]$, the following upper bound estimate holds:
\begin{equation}\label{eq33}
\begin{array}{c}
\left\| {\tilde \kappa \left( t \right)} \right\| = {e^{ - \gamma \int\limits_{{t_0}}^t {{\cal M}_\kappa ^2\left( \tau  \right)d\tau } }}\left\| {\tilde \kappa \left( {{t_0}} \right)} \right\| \le \\ \le \! {e^{ - \gamma \int\limits_{{t_e}}^t {{\cal M}_\kappa ^2\left( \tau  \right)d\tau } }}\left\| {\tilde \kappa \left( {{t_0}} \right)} \right\| \le {e^{ - \gamma \underline {{\cal M}_\kappa ^2} \left( {t - {t_e}} \right)}}\left\| {\tilde \kappa \left( {{t_0}} \right)} \right\|.
\end{array}
\end{equation}

The exponential convergence of the error $\tilde x\left( t \right)$ to zero follows jointly from \eqref{eq32} and \eqref{eq33} under assumption 1 and owing to the fact that the matrices ${A_K}{\rm{,\;}}{A_f}$ are Hurwitz ones.

Therefore, the developed adaptive observer of the system \eqref{eq3} states consists of filtering \eqref{eq13}, procedure of dimensionality reduction of the unknown parameter vector \eqref{eq14}, procedure of dynamic regressor extension and mixing \eqref{eq17}, \eqref{eq18}, procedures to transform the regression equation \eqref{eq19} into \eqref{eq26}, algebraic equation to reconstruct states with the adaptative law \eqref{eq31}. In case the regressor $\overline \varphi \left( t \right)$ obtained after application of the dimensionality reduction procedure \eqref{eq14} is finitely exciting, such an observer ensures exponential convergence of the estimates of the system \eqref{eq3} unmeasured states to their true values. The applicability conditions of the proposed observer are as follows: (\emph{i}) the assumptions 1-4 are met, (\emph{ii}) the dimensionality reduction of the unknown parameters vector is possible in such a way that $\overline \varphi  \in {\rm{FE}}$ and the requirements \eqref{eq15} are met, (\emph{iii}) there exist the mappings \eqref{eq22}-\eqref{eq25} that allow one to transform the regression equation \eqref{eq19} into a new one \eqref{eq26}. These assumptions/hypotheses are not restrictive and usually met for mathematical models of physical systems (see \cite{b16} to find a reference to an example).

\section{Numerical Experiments}
The following linear system from the experimental section of \cite{b14} has been considered:
\begin{equation}\label{eq34}
\begin{gathered}
  \dot x = {\begin{bmatrix}
  0&{{\theta _1} + {\theta _2}}&0 \\ 
  { - {\theta _2}}&0&{{\theta _2}} \\ 
  0&{ - {\theta _3}}&0 
\end{bmatrix}} x + {\begin{bmatrix}
  0 \\ 
  0 \\ 
  {{\theta _3}} 
\end{bmatrix}} u+ {\begin{bmatrix}
{{\theta _1}{\theta _2}}\\
0\\
0
\end{bmatrix}}\delta, \hfill \\
  y = {\begin{bmatrix}
  0&0&1 
\end{bmatrix}} x. \hfill \\ 
\end{gathered}
\end{equation}

For all $\theta  \in {D_\theta } = \left\{ {\theta  \in {\mathbb{R}^{{n_\theta }}}{\rm{:\;}}{\theta _2} \ne 0,{\rm{\;}}{\theta _3} \ne 0} \right\}$ the matrix ${T_I}\left( \theta  \right)$ was written as:
\begin{equation}\label{eq35}
    {T_I}\left( \theta  \right) = {\begin{bmatrix}
{ - {\textstyle{{{\theta _1} + {\theta _2}} \over {{\theta _3}}}}}&0&{{\textstyle{1 \over {{\theta _2}{\theta _3}}}}}\\
0&{ - {\textstyle{1 \over {{\theta _3}}}}}&0\\
1&0&0
\end{bmatrix}}{\rm{,}}
\end{equation}
or, following \eqref{eq24}:
\begin{equation*}
\begin{array}{c}      
{\cal Q}\left( \theta  \right) = {\begin{bmatrix}
{ - {\theta _2}\left( {{\theta _1} + {\theta _2}} \right)}&0&1\\
0&{ - 1}&0\\
1&0&0
\end{bmatrix}}{\rm{,\;}}\\{\cal P}\left( \theta  \right) = {\rm{diag}}\left\{ {{\theta _2}{\theta _3}{\rm{, }}{\theta _3}{\rm{, 1}}} \right\}.
\end{array}
\end{equation*}

The system \eqref{eq34} represented in the observer canonical form was described by the following parameters:
\begin{equation}\label{eq36}
\begin{array}{c}      
{\psi _a} = {\begin{bmatrix}
0\\
{ - \left( {{\theta _1} + {\theta _2} + {\theta _3}} \right){\theta _2}}\\
0
\end{bmatrix}}{\rm{,}}\\{\psi _b} = {\begin{bmatrix}
{{\theta _3}}\\
0\\
{{\theta _3}{\theta _2}\left( {{\theta _2} + {\theta _1}} \right)}
\end{bmatrix}}{\rm{,\;}}{e_n}{\psi _d} = {\begin{bmatrix}
0\\
0\\
{{\theta _1}\theta _2^2{\theta _3}}
\end{bmatrix}},
\end{array}
\end{equation}
where ${\psi _{ab}}\left( \theta  \right) = {\rm{col}}\left\{ { - \left( {{\theta _1} + {\theta _2} + {\theta _3}} \right){\theta _2}{\rm{,\;}}{\theta _3}{\rm{,\;}}{\theta _3}{\theta _2}\left( {{\theta _2} + {\theta _1}} \right)} \right\}.$

The parameters of the disturbance exosystem \eqref{eq4} was set as:
\begin{equation} \label{eq37}
{{\cal A}_\delta }\left( \rho  \right) = {\begin{bmatrix}
0&1\\
\rho &0
\end{bmatrix}}{\rm{,\;}}h_\delta ^{\rm{T}} = {\begin{bmatrix}
1&0
\end{bmatrix}}{\rm{,\;}}\rho  =  - 10,
\end{equation}
so then ${\Gamma ^{\rm{T}}} = {\begin{bmatrix}
0&\rho &0
\end{bmatrix}} $.

The filters \eqref{eq13} parameters were chosen as:
\begin{equation}\label{eq38}
{f^{\rm{T}}} = {\begin{bmatrix}
{ - 125}&{ - 75}&{ - 15}
\end{bmatrix}} {\rm{,\;}}{K^{\rm{T}}} = {\begin{bmatrix}
3&3&1
\end{bmatrix}}.
\end{equation}

Preliminary simulations revealed that the following equalities were true for the parameterization \eqref{eq9} for all $\theta  \in {D_\theta }$, all $u\left(t\right)$ and chosen parameters \eqref{eq38} of the filters \eqref{eq13}:
\begin{equation}\label{eq39}
    e_2^{\rm{T}}{\overline \varphi _e}\left( t \right) = e_8^{\rm{T}}{\overline \varphi _e}\left( t \right){\rm{,\;}}e_6^{\rm{T}}{\overline \varphi _e}\left( t \right) = e_{20}^{\rm{T}}{\overline \varphi _e}\left( t \right){\rm{,}}
\end{equation}
which, jointly with \eqref{eq36} and \eqref{eq37}, motivated us to choose ${{\cal D}_\eta }$ and ${{\cal L}_\eta }$ in such a way that $\eta \left( \psi  \right)$ from hypothesis 1 took the following form\footnote{These matrices are not presented in an explicit form due to space limitation.}:
\begin{equation*} 
    \eta \left( \psi  \right) = {{\begin{bmatrix}
{{\psi _{a2}} + \rho }&{{\psi _{b1}}}&{{\psi _{b3}} - {\psi _{b1}}\rho }&{ - {\psi _{a2}}\rho }&{ - {\psi _{b3}}\rho }
\end{bmatrix}}^{\rm{T}}}.
\end{equation*}

The inverse function to recalculate the parameters $\eta \left( \psi  \right)$ into $\psi$ was defined as follows (for more details please see Supplementary Material \cite{b16}):
\begin{equation}\label{eq41}
\begin{array}{l}
{\psi _1} = {\psi _3} = {\psi _5} = {\psi _7} = {\psi _9} = 0,\\
{\psi _2} = {\eta _1} + {\textstyle{{{\eta _4}{\eta _3} - {\eta _1}{\eta _5}} \over {{\eta _5} + {\eta _4}{\eta _2}}}},\\
{\psi _4} = {\eta _2},\\
{\psi _6} = {\textstyle{{{\eta _5}\left( {{\eta _5} + {\eta _4}{\eta _2}} \right)} \over {{\eta _4}{\eta _3} - {\eta _1}{\eta _5}}}},\\
{\psi _8} = {\textstyle{{{\eta _4}{\eta _3} - {\eta _1}{\eta _5}} \over { - \left( {{\eta _5} + {\eta _4}{\eta _2}} \right)}}}.
\end{array}
\end{equation}
or, following \eqref{eq22}:
\begin{equation*}
\begin{array}{l}
{{\cal S}_\psi }\left( \eta  \right) = {\rm{col}}\left\{ {0,\left( {{\eta _5} + {\eta _4}{\eta _2}} \right){\eta _1} + {\eta _4}{\eta _3} - {\eta _1}{\eta _5}{\rm{,\;0}}{\rm{,\;}}{\eta _2}{\rm{,}}} \right.\hfill\\\hfill
\left. {{\rm{ 0}}{\rm{,\;}}{\eta _5}\left( {{\eta _5} + {\eta _4}{\eta _2}} \right){\rm{,\;0}}{\rm{,\;}}{\eta _4}{\eta _3} - {\eta _1}{\eta _5}{\rm{,\;0}}} \right\}{\rm{,}}\\
{{\cal G}_\psi }\left( \eta  \right) = {\rm{diag}}\left\{ {1,{\rm{\;}}{\eta _5} + {\eta _4}{\eta _2}{\rm{,\;1}}{\rm{,\;}}1,{\rm{\;1}}{\rm{,\;}}} \right.\hfill\\\hfill
\left. {{\eta _4}{\eta _3} - {\eta _1}{\eta _5}{\rm{,\;1}}{\rm{,\;}} - \left( {{\eta _5} + {\eta _4}{\eta _2}} \right){\rm{,\;1}}} \right\}.
\end{array}
\end{equation*}

In its turn, the inverse function to recalculate the parameters ${\psi _{ab}}$ into $\theta$ was defined as:
\begin{equation}\label{eq42}
\theta  = {\cal F}\left( {{\psi _{ab}}} \right) = {\begin{bmatrix}
{{\textstyle{{\psi _{2ab}^4{\psi _{3ab}} - {\psi _{2ab}}{{\left( {{\psi _{1ab}}{\psi _{2ab}} + {\psi _{3ab}}} \right)}^2}} \over { - \psi _{2ab}^3\left( {{\psi _{1ab}}{\psi _{2ab}} + {\psi _{3ab}}} \right)}}}}\\
{{\textstyle{{{\psi _{1ab}}{\psi _{2ab}} + {\psi _{3ab}}} \over { - \psi _{2ab}^2}}}}\\
{{\psi _{2ab}}}
\end{bmatrix}},
\end{equation}
or, following \eqref{eq23}:
\begin{equation*}
    \begin{array}{c}
{\cal S}_{\theta}\left( {{\psi _{ab}}} \right) = {\begin{bmatrix}
{\psi _{2ab}^4{\psi _{3ab}} - {\psi _{2ab}}{{\left( {{\psi _{1ab}}{\psi _{2ab}} + {\psi _{3ab}}} \right)}^2}}\\
{{\psi _{1ab}}{\psi _{2ab}} + {\psi _{3ab}}}\\
{{\psi _{2ab}}{\psi _{1ab}}}
\end{bmatrix}} {\rm{,}}\\
{\cal G}_{\theta}\left( {{\psi _{ab}}} \right) \!=\! {\rm{diag}}\!\left\{ { - \psi _{2ab}^3\left( {{\psi _{1ab}}{\psi _{2ab}} \!+\! {\psi _{3ab}}} \right)\!{\rm{,}} - \psi _{2ab}^2{\rm{,}}{\psi _{1ab}}} \right\}\!.
\end{array}
\end{equation*}

Finally, the matrix ${{\cal O}_\Gamma }\left( \Gamma  \right)$ was defined as:
\begin{equation}\label{eq43}
{{\cal O}_\Gamma }\left( \Gamma  \right) = {\begin{bmatrix}
{ - {f_1}}&{\rho  - {f_2}}&{ - {f_3}}\\
0&{ - {f_1} - {f_3}\rho }&{\rho  - {f_2}}\\
0&{ - \rho \left( {{f_2} - \rho } \right)}&{ - {f_1} - {f_3}\rho }
\end{bmatrix}}.
\end{equation}

Having used \eqref{eq35}, \eqref{eq41}-\eqref{eq43}, the functions from lemma were formed:
\begin{equation}\label{eq44}
\begin{array}{c}
{{\cal T}_{\cal Q}}\left( {{{\overline \Xi }_{\cal Q}}\left( {{{\cal M}_\theta }} \right){{\cal Y}_\theta }} \right) = \\ = {\begin{bmatrix}
{ - {{\cal Y}_{\theta 2}}\left( {{{\cal Y}_{\theta 1}} + {{\cal Y}_{\theta 2}}} \right)}&0&{{\cal M}_\theta ^2}\\
0&{ - {{\cal M}_\theta }}&0\\
{{{\cal M}_\theta }}&0&0
\end{bmatrix}} {\rm{,}}\\
{{\cal T}_{\cal P}}\left( {{{\overline \Xi }_{\cal P}}\left( {{{\cal M}_\theta }} \right){{\cal Y}_\theta }} \right) = {\rm{diag}}\left\{ {{{\cal Y}_{\theta 2}}{{\cal Y}_{\theta 3}}{\rm{,\;}}{{\cal Y}_{\theta 3}}{\rm{,\;}}{{\cal M}_\theta }} \right\}{\rm{,}}
\end{array}
\end{equation}
\begin{equation}\label{eq45}
\begin{array}{l}
{{\cal T}_{{{\cal S}_\psi }}}\left( {{{\overline \Xi }_{{{\cal S}_\psi }}}\left( \Delta  \right){\cal Y}} \right){\rm{ = }}\\
{\rm{ = col}}\left\{ {0,\;\left( {{{\cal Y}_5}\Delta  + {{\cal Y}_4}{{\cal Y}_2}} \right){{\cal Y}_1} + \Delta \left( {{{\cal Y}_4}{{\cal Y}_3} - {{\cal Y}_1}{{\cal Y}_5}} \right){\rm{,}}} \right.\\
\hfill{\rm{0}}{\rm{,\;}}\left. {{{\cal Y}_2}{\rm{,\;0}}{\rm{,\;}}{{\cal Y}_5}\left( {\Delta {{\cal Y}_5} + {{\cal Y}_4}{{\cal Y}_2}} \right){\rm{,\;0}}{\rm{,\;}}{{\cal Y}_4}{{\cal Y}_3} - {{\cal Y}_1}{{\cal Y}_5}{\rm{,0}}} \right\}\\
{{\cal T}_{{{\cal G}_\psi }}}\left( {{{\overline \Xi }_{{{\cal G}_\psi }}}\left( \Delta  \right){\cal Y}} \right) = \\
 = {\rm{diag}}\left\{ {\Delta {\rm{,\;}}{\Delta ^2}{{\cal Y}_5} + \Delta {{\cal Y}_4}{{\cal Y}_2}{\rm{,\;}}\Delta {\rm{,\;}}\Delta {\rm{,\;}}\Delta {\rm{,}}} \right.\\
\hfill\left. {\Delta \left( {{{\cal Y}_4}{{\cal Y}_3} - {{\cal Y}_1}{{\cal Y}_5}} \right){\rm{,\;}}\Delta {\rm{,\;}} - \left( {\Delta {{\cal Y}_5} + {{\cal Y}_4}{{\cal Y}_2}} \right){\rm{,}}\Delta } \right\}
\end{array}
\end{equation}
\begin{equation}\label{eq46}
\begin{array}{c}
{{\cal T}_{{{\cal S}_\theta }}}\left( {{{\overline \Xi }_{{{\cal S}_\theta }}}\left( {{{\cal M}_\psi }} \right){{\cal Y}_{ab}}} \right) = \\
 = {\begin{bmatrix}
{{\cal Y}_{2ab}^4{{\cal Y}_{3ab}} - {{\cal Y}_{2ab}}{{\left( {{{\cal Y}_{1ab}}{{\cal Y}_{2ab}} + {{\cal M}_\psi }{{\cal Y}_{3ab}}} \right)}^2}}\\
{{{\cal Y}_{1ab}}{{\cal Y}_{2ab}} + {{\cal M}_\psi }{{\cal Y}_{3ab}}}\\
{{{\cal Y}_{2ab}}{{\cal Y}_{1ab}}}
\end{bmatrix}}{\rm{,}}\\
{{\cal T}_{{{\cal G}_\theta }}}\left( {{{\overline \Xi }_{{{\cal G}_\theta }}}\left( {{{\cal M}_\psi }} \right){{\cal Y}_{ab}}} \right) = \\
 = {\rm{diag}}\left\{ \begin{array}{c}
 - {\cal Y}_{2ab}^3\left( {{{\cal Y}_{1ab}}{{\cal Y}_{2ab}} + {{\cal M}_\psi }{{\cal Y}_{3ab}}} \right){\rm{,}}\\
{\rm{ }} - {\cal Y}_{2ab}^2{\rm{,\;}}{{\cal M}_\psi }{{\cal Y}_{1ab}}
\end{array} \right\}{\rm{,}}
\end{array}
\end{equation}
\begin{equation}\label{eq47}
\begin{array}{c}
{{\cal T}_{{{\cal O}_\Gamma }}}\left( {{{\overline \Xi }_{{{\cal O}_\Gamma }}}\left( {{{\cal M}_\psi }} \right){{\cal Y}_\Gamma }} \right) = \\ = {{\left[{\begin{smallmatrix}
{ - {{\cal M}_\psi }{f_1}}&{{{\cal Y}_{2\Gamma }} - {{\cal M}_\psi }{f_2}}&{ - {{\cal M}_\psi }{f_3}}\\
0&{ - {{\cal M}_\psi }{f_1} - {f_3}{{\cal Y}_{2\Gamma }}}&{{{\cal Y}_{2\Gamma }} - {{\cal M}_\psi }{f_2}}\\
0&{ - {{\cal Y}_{2\Gamma }}\left( {{{\cal M}_\psi }{f_2} - {{\cal Y}_{2\Gamma }}} \right)}&{ - {\cal M}_\psi ^2{f_1} - {{\cal M}_\psi }{f_3}{{\cal Y}_{2\Gamma }}}
\end{smallmatrix}}\right]{\rm{,}}}}\\
{\Pi _{{{\cal O}_\Gamma }}}\left( {{{\cal M}_\psi }} \right) = {\rm{diag}}\left\{ {{{\cal M}_\psi }{\rm{,\;}}{{\cal M}_\psi }{\rm{,\;}}{\cal M}_\psi ^2} \right\},
\end{array}
\end{equation}
and, consequently, we obtained equations from Lemma to calculate the signals ${{\cal Y}_\kappa }\left( t \right)$ and ${{\cal M}_\kappa }\left( t \right)$.

The parameters of system \eqref{eq34}, filters \eqref{eq17}, mixing procedure \eqref{eq18}, control law and adaptive law \eqref{eq31} were picked as:
\begin{equation*}
\begin{array}{l}
{\theta _1} = 1,{\rm{\;}}{\theta _2} = 1,{\rm{\;}}{\theta _3} =  - 1,{\rm{\;}}\sigma  =  - 1,{\rm{\;}}{t_{\epsilon}} = 25,{\rm{\;}}\\k\left( t \right) = {\textstyle{1 \over {{\rm{det}}\left\{ {\varphi \left( t \right)} \right\} + {{10}^{ - 19}}}}}{\rm{,\;}}\gamma  = {\rm{1}}{\rm{,\;}}\hat \kappa_0 = 10{\rm{rand}}\left( {27,1} \right){\rm{,\;}}\\u =  - 75\left( {2.{\rm{5}}h\left( {t - {t_{\epsilon}}} \right){\rm{sin}}\left( {10t} \right){e^{ - 1\left( {t - {t_{\epsilon}}} \right)}} + 100 - y} \right){\rm{,}}
\end{array}
\end{equation*}
where $h\left( {t - {t_{\epsilon}}} \right)$ stands for a Heaviside step function, ${\rm{rand}}\left( {27,1} \right)$ is a MatLab function that returns a 27-by-1 matrix containing pseudorandom values drawn from the standard uniform distribution over an open interval $\left(0,\;1\right)$.

The observer \eqref{eq20a}, \eqref{eq20b}  was also implemented with $\gamma  =\linebreak = 1,{\rm{\;}}\hat \eta \left( {{t_0}} \right) = 10{\rm{rand}}\left( {5,{\rm{ 1}}} \right)$ for the comparison purposes.

Figure 1 presents the difference $\overline q\left( t \right) - {\overline \varphi ^{\rm{T}}}\left( t \right)\eta \left( \psi  \right)$, as well as the value of ${\lambda _{{\rm{min}}}}\left\{ {\int\limits_t^{t + T} {\overline \varphi \left( \tau  \right){{\overline \varphi }^{\rm{T}}}\left( \tau  \right)d\tau } } \right\}$ for $T = 1$.

The transients shown in Fig. 1 validated the conclusions made in propositions 1 and 2. For all $t \ge {t_{\epsilon}} = 25$ it held that $\overline q\left( t \right) = {\overline \varphi ^{\rm{T}}}\left( t \right)\eta \left( \psi  \right)$. The transient curve of the minimum eigenvalue of the integral showed that the injection of $2.{\rm{5sin}}\left( {10t} \right){e^{ - 1\left( {t - {t_{\epsilon}}} \right)}}$ into the control signal allowed one to meet the condition $\overline \varphi  \in {\rm{FE}}$.
\begin{figure}[htbp]
\centerline{\includegraphics[scale=0.55]{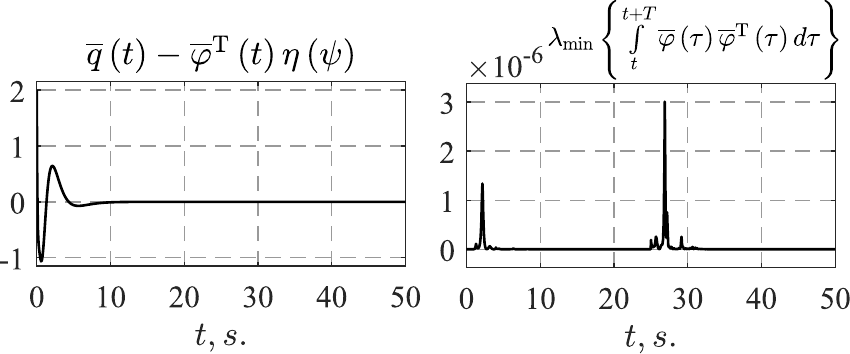}}
	\caption{Comparison of $ \overline q\left( t \right) = {f^{\rm{T}}}F\left( t \right) + y\left( t \right) - C_0^{\rm{T}}z\left( t \right)$ with ${\overline \varphi ^{\rm{T}}}\left( t \right)\eta \left( \psi  \right)$ and transient response of ${\lambda _{{\rm{min}}}}\left\{ {\int\limits_t^{t + T} {\overline \varphi \left( \tau  \right){{\overline \varphi }^{\rm{T}}}\left( \tau  \right)d\tau } } \right\}$.}
	\label{fig1}
\end{figure}

Figure 2 shows behavior of parametric errors ${\tilde \psi _a}\left( t \right) = \linebreak = {\hat \psi _a}\left( t \right) - {\psi _a}\left( \theta  \right){\rm{,\;}}{\tilde \psi _b}\left( t \right) = {\hat \psi _b}\left( t \right) - {\psi _b}\left( \theta  \right){\rm{,}}$ ${\tilde {\cal O}_\Gamma }\left( t \right) = \linebreak = {\hat {\cal O}_\Gamma }\left( t \right) - {{\cal O}_\Gamma }\left( \Gamma  \right){\rm{,\;}}{\tilde T_I}\left( t \right) = {\hat T_I}\left( t \right) - {T_I}\left( \theta  \right)$ when the proposed observer \eqref{eq31} was applied.
\begin{figure}[htbp]
\centerline{\includegraphics[scale=0.55]{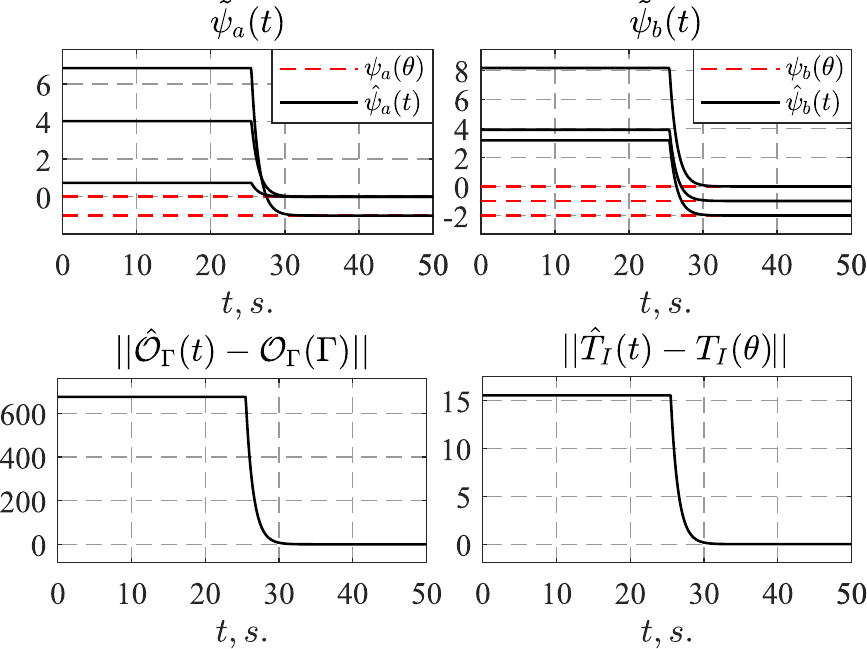}}
	\caption{Transient behavior of parametric errors.}
	\label{fig2}
\end{figure}

Figure 3 depicts the behavior of the state reconstruction error $\tilde x\left( t \right) = \hat x\left( t \right) - x\left( t \right)$ for the observers \eqref{eq31} and (19).

\begin{figure}[htbp]
\centerline{\includegraphics[scale=0.55]{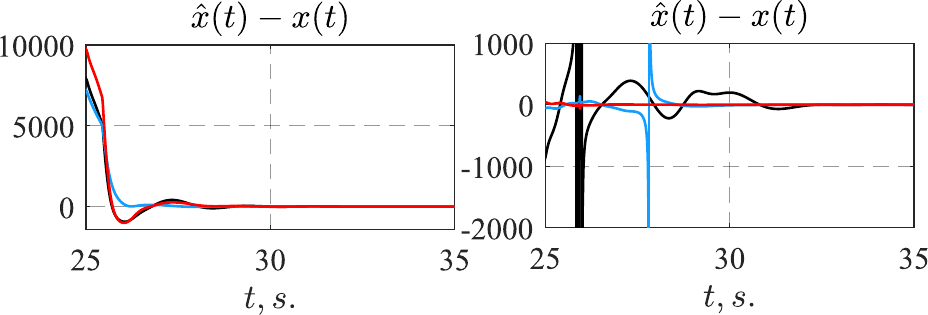}}
	\caption{Transient response of $\tilde x\left( t \right) = \hat x\left( t \right) - x\left( t \right)$ for \eqref{eq31} and (19).}
	\label{fig3}
\end{figure}

The simulation results validated the conclusions made within the theoretical analysis. The proposed adaptive observer \eqref{eq31} reconstructed the physical state of the system affected by an external disturbance generated by an exosystem with fully uncertain constant parameters. In contrast to the certainty-equivalence-based observer (19), the obtained state estimate did not contain discontinuities, and the observer \eqref{eq31} itself did not require to meet the Lipschitz condition \eqref{eq21}.

\section{Conclusion}

For the uncertain linear systems with overparameterization and unknown additive perturbations generated by an exosystem with fully uncertain constant parameters, a novel adaptive observer is proposed, which ensures exponential convergence of the physical states reconstruction error to zero when the condition of finite excitation is satisfied.

Unlike the solution from \cite{b12}, the proposed observer allows one to reconstruct the physical state $x\left( t \right)$ of the system \eqref{eq3} rather than the virtual one $\xi \left( t \right)$ of the observer canonical form \eqref{eq7}. In contrast to the earlier result \cite{b13, b14}, the adaptive observer proposed in this study is applicable to systems with unknown additive perturbations generated by an exosystem with fully uncertain constant parameters (i.e. the condition that only the initial conditions of \eqref{eq4} are unknown is removed). To the best of authors’ knowledge, the proposed observer is the first solution of physical state reconstruction problem for fully unknown linear systems with overparameterization and unknown additive perturbations.

\end{document}